# Polaritonic-to-Plasmonic Transition in Optically Resonant Bismuth Nanospheres for High-Contrast Switchable Ultraviolet Meta-Filters

Alexander Cuadrado[1], Johann Toudert[1,*], Rosalia Serna[1]

[1]*Laser Processing Group, Instituto de Óptica, CSIC, Serrano 121, 28006 Madrid, Spain*
*\* corresponding author: johann.toudert@gmail.com*

**Abstract** In the quest aimed at unveiling alternative plasmonic elements overcoming noble metals for selected applications in photonics, we investigate by numerical simulations the near ultraviolet-to-near infrared optical response of solid and liquid Bi nanospheres embedded in a dielectric matrix. We also determine the resulting transmission contrast upon reversible solid-liquid phase transition to evaluate their potential for switchable optical filtering. The optical response of the solid (liquid) Bi nanospheres is ruled by localized polaritonic (plasmonic) resonances tunable by controlling the diameter. For a selected diameter between 20 nm and 50 nm, both solid and liquid nanospheres present a dipolar resonance inducing a strong peak extinction in the near ultraviolet, however at different photon energies. This enables a high transmission contrast at selected near ultraviolet photon energies. It is estimated that a two-dimensional assembly of 30 nm solid Bi nanospheres with a surface coverage of 32% will almost totally extinct (transmission of 2%) a near ultraviolet 3.45 eV (359 nm) light beam, whereas upon phase transition the resulting liquid Bi nanospheres will show a transmission of 30%. This work appeals to the fabrication of locally reconfigurable optical metamaterials for integrated switchable near ultraviolet optics.

**Keywords** Bismuth, Plasmon Resonance, Phase Transition, Reconfigurable Metamaterials, Ultraviolet.

# 1. Introduction

The forthcoming generation of optical metamaterials will operate in the near ultraviolet (UV)-to-near infrared (IR) region and be locally reconfigurable: they will display localized optical resonances in this spectral region that will be switchable locally, down to the nanoscale. Such features are required for applications in integrated photonics such as switchable optical filtering, tunable refraction of light or tunable lasing.[1-4]

Switchable localized optical resonances have been already demonstrated in reconfigurable metamaterials consisting of noble metal nanostructures (Au or Ag nanospheres, nanorods, nanodiscs) embedded in/deposited on a matrix/subtrate with reversibly tunable optical properties, consisting for instance of a phase-transition material.[5-13] In these metamaterials, the resonances are induced by the plasmonic response of the noble metal nanostructures and the switchable response (modulation in the peak photon energy, amplitude and width of the resonances) is based on the change in the dielectric function of the tunable matrix/substrate upon its phase transition triggered by an external stimulus (light, voltage, heat...). In other words, these metamaterials are based on two *separated* types of functional elements: *resonators* (the noble metal nanostructures) and *switch* (the tunable matrix/subtrate). However, in order to achieve a *local reconfigurability* of the metamaterial down to the nanoscale demanded for an ultimate control in the light-metamaterial interaction, it will be desirable that these *two functionalities (resonator and switch)* could be *merged in a single nanoscale element,*[14] *with a markedly subwavelength size*. This requires nanostructures presenting a strong sensitivity of their dielectric function to external stimuli, that could therefore be supported on/embedded in usual (non-tunable) dielectric substrates/matrices. Noble metal nanostructures do not fulfil this requirement as their dielectric function is very little sensitive to external stimuli.

Another important limitation of noble metal nanostructures embedded in usual dielectric matrices (with a moderate to high refractive index, n > 1.5) is that they can be easily designed for supporting localized optical resonances in the visible and near-infrared,[15-17] *but not in the near UV*.

Therefore, achieving locally reconfigurable metamaterials based on embedded nano-resonators and operating in the near UV demands the development of non-conventional plasmonic nanostructures, beyond noble metals.[18-19] Several metals show excellent plasmonic performance in the UV region,[20] such as Al,[21-22] In[22] or Rh.[23] However the dielectric functions of these metals are hardly sentitive to external stimuli. In contrast, plasmonic nanostructures based on phase-transition materials are more suitable candidates,[24-29] as they are able to include the *resonator* and *switch* functionalities. A special attention has been focussed on the single-element material Ga, which is an excellent plasmonic candidate for optically induced switching at low power density, due to its near-room temperature bulk melting point (30ºC, 303 K).[24] However, Ga nanostructures present substantially lower melting and solidification temperatures than bulk Ga, i.e. much below room temperature.[25] Switching based on Ga nanostructures thus requires operation with an active cooling system. In this context, a promising single-element material for switching applications is Bi. The melting and solidification temperatures of embedded Bi nanostructures are higher than room temperature[26] thus making them suitable for switching without need of active cooling, and using instead active heating. However, their potential for supporting *strong and switchable resonances in the near UV* when *markedly subwavelength in size* and *embedded in a usual dielectric matrix* has to be explored.

## 2. Solid-Liquid Transition of Bismuth for Optical Switching: From Thin Films to Optically Resonant Nanostructures

Bi, as single-element material, presents a solid-liquid transition around 270ºC (543K, for both the bulk material and embedded nanostructures)[26] with a significant contrast in the near UV-to-near IR between the dielectric function $\varepsilon$ of the solid and liquid phase. This can be clearly seen on figure 1a, that shows the spectra of the refractive index n and extinction coefficient k (fulfiling the relation $\varepsilon = (n+jk)^2$) of bulk solid[17] and liquid Bi.[30]

As a straightforward consequence, Bi thin films could be considered suitable candidates for the fabrication of *switchable* optical filters based on the solid-liquid transition of Bi. Figure 1b (top panel) indeed shows significant differences between the near UV-to-near IR transmittance spectra of a 10 nm-thick solid Bi and liquid Bi films. At photon energies below (above) 1.8 eV, the transmittance of the liquid film is higher (lower) than that of the solid film. As seen in the bottom panel of figure 1b, transmittance contrast (transmittance of the liquid film – transmittance of the solid film) values between -15% and 15% are obtained upon adjusting the film thickness. The highest contrast values are obtained for the 5 nm film, but at the cost of high transmittance values for both the solid and liquid films. Since a perfect switchable optical filter should present one state with a 0% transmittance and the other with a 100% transmittance, and thus a 100% transmittance contrast, the performance of Bi thin films as switchable optical filters is rather poor.

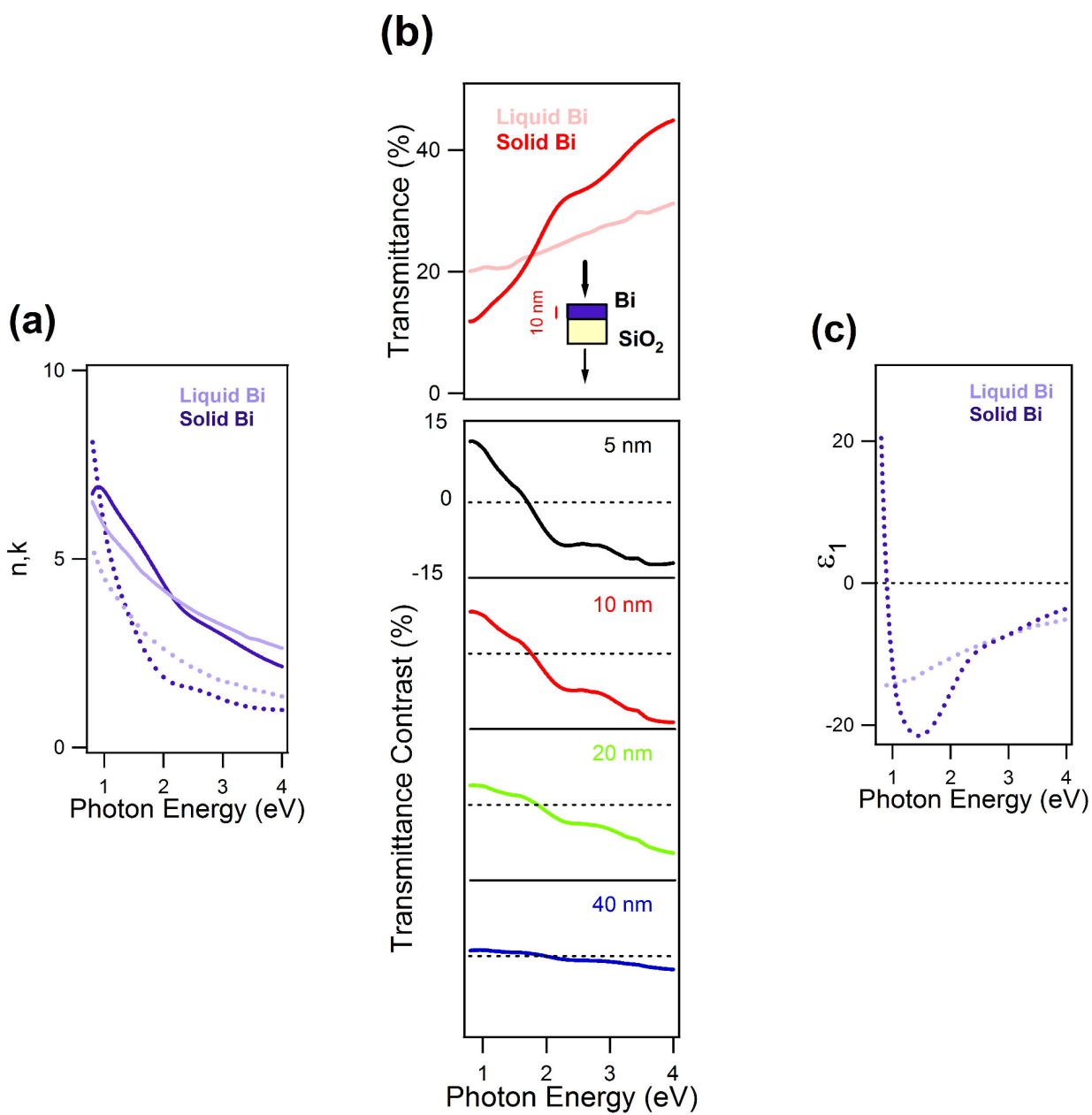

**Figure 1:** (a) Near UV-to-near IR spectra of the refractive index n (dotted lines) and extinction coefficient k (full lines) of solid (dark line) and liquid (clear line) Bi, derived from refs 17 and 30, respectively. (b) Top panel: Calculated near UV-to-near IR transmittance spectra of a solid and liquid 10 nm Bi film on a transparent $SiO_2$ substrate; Bottom panel: Transmittance contrast (transmittance of the liquid film – transmittance of the solid film) for Bi films of different thicknesses (5 nm to 40 nm). The spectra are offset vertically for the sake of clarity (offset = multiples of 30%). (c) Near UV-to-near IR spectra of the real part $\varepsilon_1$ of the dielectric function of solid and liquid Bi, taken from refs 17 and 30, respectively.

An attractive alternative is to consider Bi nanostructures embedded in a robust dielectric matrix that will encapsulate them and protect them from degradation. This matrix acts as a mould in which the Bi nanostructures can melt and solidify reversibly. Therefore their size and shape are not significantly affected by the phase transition (the solid-liquid transition induces a small 3.4% volume contraction). Thus, the key variable behind the *switchable* optical behaviour of embedded Bi nanostructures is their dielectric function $\varepsilon = \varepsilon_1 + j\varepsilon_2$, that changes significantly in the near UV-to-near IR spectral region upon solid-liquid transition. Despite the differences between their dielectric functions, both solid and liquid Bi present strongly negative $\varepsilon_1$ values in the near UV-to-near IR spectral region (as shown on figure 1c), which is a requirement for achieving localized optical resonances in this spectral region for the embedded Bi nanostructures. Such *optical resonances*[17,26] have been reported for both solid and liquid Bi nanostructures embedded in a dielectric matrix.

At this point it is interesting to note that the negative $\varepsilon_1$ values in solid Bi can be interpreted as a consequence of the excitation of interband transitions with high oscillator strength in the near IR,[31] while in liquid Bi they have mainly a (free-electron) Drude origin,[29-30] as in the case of (solid) noble metals. Therefore Bi nanostructures in the solid and liquid state show *distinct* localized optical resonances that have an interband polaritonic and plasmonic origin, respectively.[26]

Summarizing, Bi nanostructures embedded in a dielectric matrix are suitable for *combining resonator and switch functionalities up to the near UV region*. We have already shown experimentally that Bi nanostructures embedded in a robust dielectric matrix are suitable for fabricating switchable resonant optical filters,[26] the resonance energy being switched reversibly at the solid-liquid transition of Bi.[31] In these works, the nanostructures were a few nanometers in size and presented a low optical absorption efficiency. Their localized optical resonances were well accounted for by a dipolar quasi-static model.

In contrast, the optical response of Bi nanostructures of larger sizes, for which the quasi-static dipolar model should not be valid anymore and for which scattering should become relevant, remains little explored so far, even in the simplest case of nanospheres. Up to date there are only two recent reports,[32-33] where the optical response of solid Bi nanospheres have been studied. In ref. 32, the extinction of solid Bi nanospheres 60 nm in diameter in water (n ~ 1.35) was calculated using the Mie theory and compared with experimental data. In ref. 33, the extinction of solid Bi nanospheres 20 to 200 nm in diameter in vacuum (n = 1) was calculated using the Mie theory. To our knowledge, the optical response of liquid Bi nanospheres has not been measured nor calculated. By analogy with the size-dependent optical response of spherical noble metal nanospheres,[15,34] one can expect that the absorption, scattering and extinction efficiencies of spherical Bi nanospheres can be tuned as a function of their size due to purely classical electromagnetic effects. This has very interesting consequences for applications since the corresponding liquid-solid contrast spectra should be also tunable as a function of the nanosphere size, and may show enhanced values in relation with the optical resonances in the nanosphere.

Therefore in the following sections, we *investigate by a finite element method* the absorption, scattering and extinction spectra of solid and liquid Bi nanospheres embedded in a usual dielectric matrix for a wide range of diameters from 20 to 300 nm. We also calculate these spectra in the quasi-static dipolar limit (diameter → 0) for the sake of comparison. Furthermore we evaluate the transmission/extinction contrast between their response in the solid and liquid state. We discuss the usefulness of simple metamaterials based on Bi nanospheres compared to Bi thin films for the development of switchable optical filters operating in the near UV spectral region.

## 3. Exploring the Near UV-to-Near IR Optical Response of Solid and Liquid Bi Nanospheres Embedded in a Dielectric Matrix

### 3.1. Finite Element Method

Finite element calculations have been performed using the Comsol Multiphysics software.[35-39] The classical Maxwell equations have been solved for a Bi nanosphere embedded in a homogeneous dielectric matrix, excited by a plane monochromatic electromagnetic wave (electric field $E_0$). The model structure is shown in figure 2a (top panel). The Bi nanosphere is represented in blue, and is described by the photon energy - dependent complex bulk Bi dielectric function given in ref. 17 and ref. 30 for the solid and liquid, respectively (corresponding to the n and k spectra shown in figure 1a). The dielectric matrix is described by the photon energy - dependent dielectric function of a-$Al_2O_3$ given in ref. 17. This matrix, which presents moderate refractive index values (n ~ 1.65) is surrounded by a spherical perfectly matched layer (PML) concentric to the Bi nanosphere, which acts as an infinite volume without any reflections of the incident electromagnetic radiation. A high density three-dimensional meshing has been used. A tetrahedral mesh has been used in the nanosphere and the surrounding matrix, where the maximum size of this mesh is fixed to be at least 10 times smaller than the effective wavelength of incident radiation. The mesh located in the PML is based on a distribution of prismatic elements oriented along the radius of this external domain. The excellent quality of the simulation down to the smallest nanoparticle diameter D considered in this work is examplified on figure 2a (bottom panel). It shows the calculated intensity map of the electric field scattered by a 25 nm Bi liquid nanosphere in its equator plane, upon excitation at a photon energy of 3.55 eV. A well defined dipolar near-field pattern can be seen.

The absorption, scattering, and extinction efficiencies have been calculated as a function of the nanosphere diameter D at selected photon energies. These efficiencies are calculated by dividing the corresponding cross sections by the geometrical cross-section of the nanosphere, $\pi D^2/4$. The absorption efficiency $Q_{Absorption}$ has been calculated using the relation:

$$Q_{Absorption}(\text{Photon Energy}) = [\int J(\text{Photon Energy}) E(\text{Photon Energy}) dV]/[I_0 \pi D^2/4] \quad (1)$$

where J is the induced current density through the nanosphere, E is the electric field along the structure, D is the geometrical diameter, $I_0$ is incident radiation intensity, and the integration is done over all the volume elements dV contained by the spherical nanosphere. The scattering efficiency $Q_{scattering}$ is calculated as:

$$Q_{Scattering}(\text{Photon Energy}) = [\int Re(\mathbf{n} \cdot \mathbf{S}) ds]/[I_0 \pi D^2/4] \quad (2)$$

where the integration is done over the far-field sphere, **n** being the normal vector pointing outwards, **S** the Poynting vector, and ds a surface element. Finally, the (total) extinction efficiency $Q_{Extinction}$ has been calculated from:

$$Q_{Extinction}(\text{Photon Energy}) = Q_{Absorption}(\text{Photon Energy}) + Q_{Scattering}(\text{Photon Energy}) \quad (3)$$

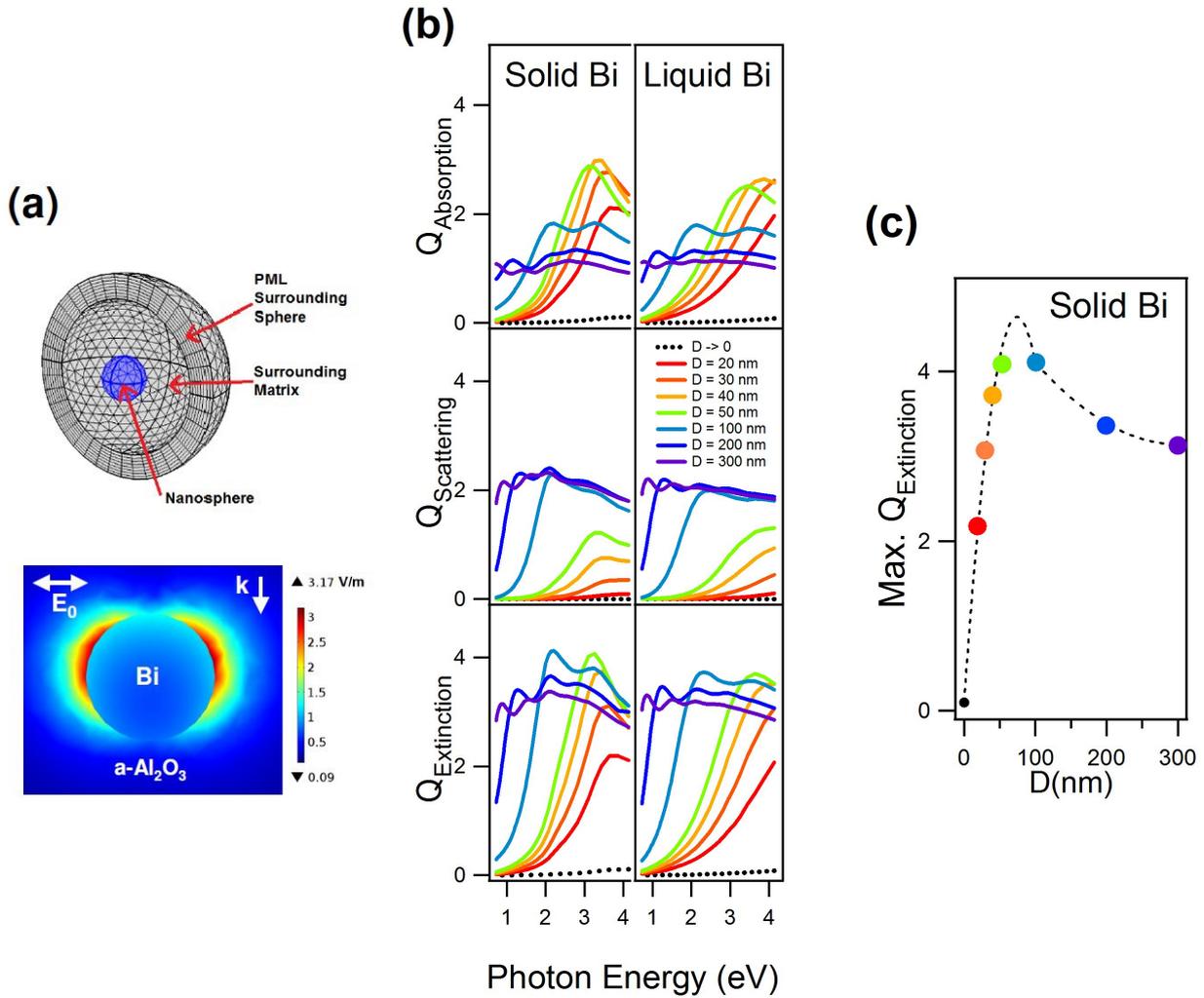

Figure 2. (a) Top panel: image showing the structure used in the model for the calculation : the nanosphere is shown in blue and the surrounding matrix and PML surrounding sphere in gray, Both present a three dimensional tetrahedral mesh. Bottom panel: Calculated intensity map of the electric field scattered by a 25 nm liquid Bi nanosphere in its equator plane, upon excitation with a plane wave at a photon energy of 3.55 eV. k is the wave vector and $E_0$ is the electric field of the incident radiation. (b) Selected simulated spectra of the absorption, scattering and extinction efficiencies ($Q_{Absorption}$, $Q_{Scattering}$ and $Q_{Extinction}$) of the solid (left column) and liquid (right column) Bi nanosphere embedded in a-$Al_2O_3$ for selected diameters D between 20 and 300 nm. The absorption and extinction spectra obtained in the quasi-static dipolar limit (D→0, no scattering) are also shown (black dotted lines). (c) Maximum of the extinction efficiency at the dipolar resonance as a function of the nanosphere diameter D (solid bismuth). The dotted curve is a guide for the eye.

## 3.2. Results

Figure 2b shows selected calculated absorption, scattering, extinction efficiency spectra for the embedded Bi nanosphere in the solid and liquid state (left and right columns, respectively), for selected diameters between 20 nm and 300 nm. The spectra show optical resonances both for the case of the solid and liquid nanospheres. The spectral features (position, width and amplitude) of these resonances depend markedly on D and on the physical state (solid or liquid) of the nanosphere, as described below.

**Multipolar resonances and scattering contribution**

For both solid and liquid nanospheres with a small diameter (D ≤ 50 nm), the spectra are dominated by a dipolar resonance in the near UV (photon energy > 3 eV). Upon increasing D above 50 nm, this resonance shifts toward lower energy while other resonances appear from the higher energy side of the spectrum and also shift toward lower energy. These resonances are attributed to multipolar orders. This behaviour is qualitatively comparable to that found for the localized plasmonic resonances of noble metal nanospheres.[15, 34, 40] Moreover, in a comparable way to Ag and Au, the contribution of scattering to extinction increases with D. For instance, the absorption efficiency/scattering efficiency ratio reaches 2 for D = 50 nm, and is as low as 0.5 for D = 300 nm.

**Bandgap-like response**

The spectra for D > 100 nm present a "bandgap-like" response, with a low (high) absorption, scattering and extinction efficiency on the low (high) energy side of the dipolar resonance. Therefore, increasing the diameter of a nanosphere makes its spectral range of high absorption, scattering and extinction efficiencies extend toward lower photon energies. As a consequence, at the largest D, the absorption, scattering and extinction efficiencies become almost photon-energy independent in the photon energy range studied in this work.

**Nanosphere Diameter for Maximum Extinction Efficiency at the Dipolar Resonance**

High extinction efficiencies are obtained for both the solid and liquid Bi nanospheres with 20 nm ≤ D ≤ 50 nm at their dipolar resonance in the near UV. These extinction efficiency values are much higher than those obtained in the quasi-static dipolar limit (D→0),[17] also shown in figure 2b for comparison (black dotted lines). This suggests that an optimum D that maximizes $Q_{Extinction}$ can be found in the 0 – 100 nm range. In order to accurately determine this optimum D for the solid Bi nanospheres, the maximum $Q_{Extinction}$ value at the dipolar resonance for several D values in the 0 – 100 nm range has been plotted in figure 2c. It can be seen that the optimum D is between 50 and 100 nm. Moreover, $Q_{ext}$ increases fast as a function of D in the 0 – 50 nm range, and then decreases more slowly. This trend is very similar to that observed for the dipolar plasmon resonances of Ag and Au nanospheres.[15] Note that the maximum values of $Q_{Extinction}$ (4.1) in the near UV is comparable to those obtained in the visible region for noble metal nanospheres (~7.5 at ~2.2 eV for Au and ~11 at ~2.95 eV for Ag in water).[15] Similar trends are observed for the liquid Bi nanospheres, however with different photon energies of the $Q_{Extinction}$ compared with the solid Bi nanospheres.

**Liquid-Solid Transmission Contrast**

For *switchable optical filtering* applications based on the embedded Bi nanospheres, it is necessary to find out the conditions for optimum extinction contrast upon phase transition. Remind that upon phase transition, the diameter of nanospheres embedded in a robust dielectric matrix will not change significantly (we legitimately neglect the 3.4% contraction upon solid-liquid transition). Therefore, we now compare the $Q_{Extinction}$ spectra of solid and liquid nanospheres of the same diameter.

As seen in figure 2b, the largest differences between the $Q_{Extinction}$ spectra of solid and liquid Bi nanospheres are obtained for 20 nm ≤ D ≤ 50 nm. This occurs in the spectral region of their dipolar resonance (near UV) that peaks at a markedly lower photon energy for a solid nanosphere than for a liquid nanosphere with the same diameter. At the largest sizes (D>100 nm), the optical spectra of the liquid and solid nanospheres are almost identical. In order to quantify these trends and to qualitatively compare these results with the transmittance contrast shown on figure 1b for the Bi films, we have calculated for each nanosphere diameter the liquid-solid "transmission contrast" defined as : - [$Q_{Extinction\ Liquid\ Bi}$ - $Q_{Extinction\ Solid\ Bi}$], which is related with the transmitted light intensity in the forward direction (i.e. not scattered nor absorbed).

As shown in figure 3a, the transmission contrast spectrum presents resonant features that can be tuned with the diameter of the nanosphere. The nanospheres with 20 nm ≤ D ≤ 50 nm yield the highest peak contrast values which are located in the near UV region. Note that these peak values are positive, while Bi thin films only allowed achieving negative transmittance contrasts in the near UV region (see figure 1b).

These results are quantified on figure 3b that shows the maximum transmission contrast and the corresponding photon energy, for the different D values. When increasing D from 20 to 50 nm, the photon energy of maximum transmission contrast decreases from 3.4 eV to 2.9 eV. A maximum contrast of 0.9 is obtained near D = 30 nm at a near UV photon energy of 3.3 eV, in relation with the lower peak energy of the resonance of the solid Bi nanosphere when compared with that of the liquid nanosphere (as seen on figure 2b).

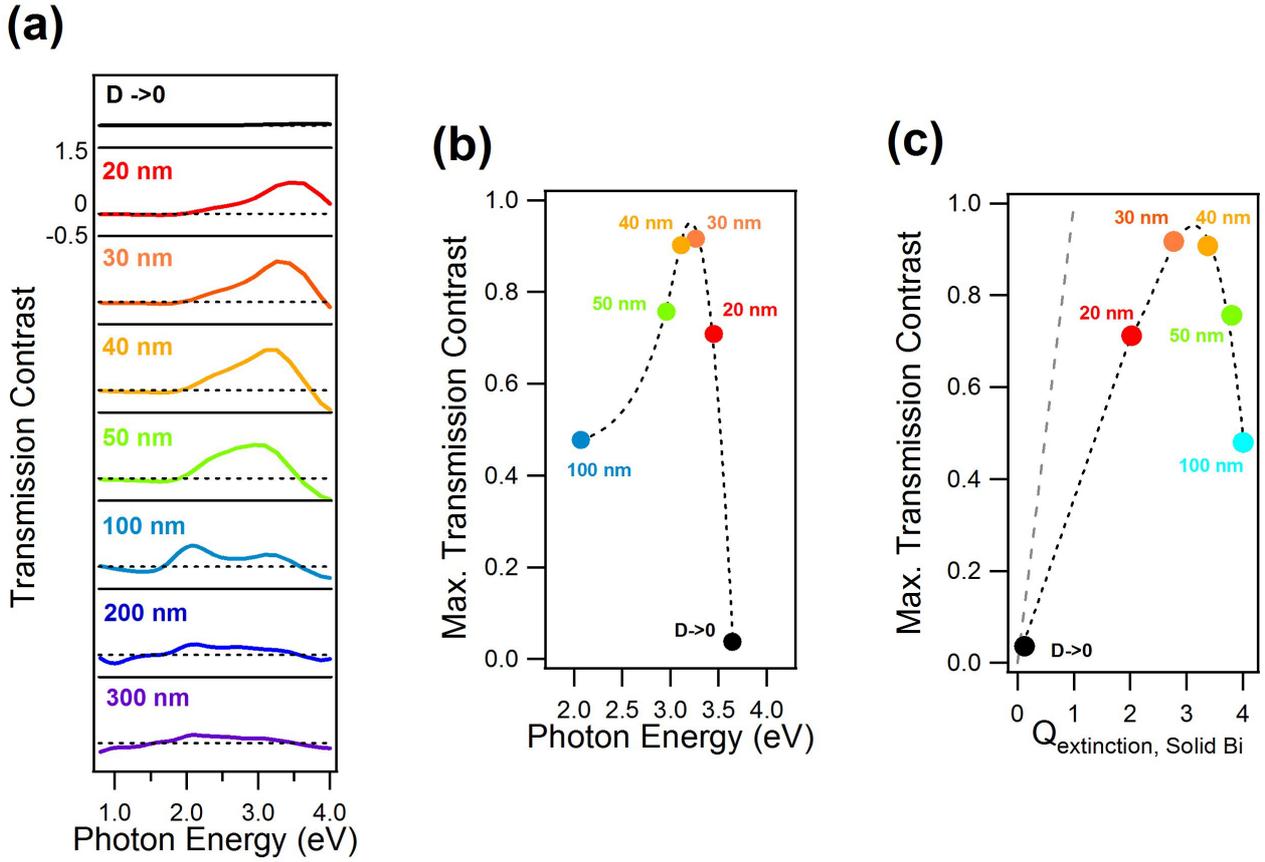

Figure 3 (a) "Transmision contrast" (- [$Q_{Extinction\ Liquid\ Bi}$ - $Q_{Extinction\ Solid\ Bi}$]) spectra calculated for a Bi nanosphere, as a function of its diameter D up to 300 nm. The spectra are offset vertically for the sake of clarity (offset = multiples of 4) (b) Maximum transmission contrast and corresponding photon energy, for D between 0 and 100 nm. (c) Maximum transmission contrast and extinction efficiency of the solid Bi nanosphere $Q_{Extinction\ Solid\ Bi}$ (at the same photon energy), for D between 0 and 100 nm. The dashed line represents the case where the maximum transmission contrast would be equal to $Q_{Extinction\ Solid\ Bi}$ (case of a "perfect switch"). The dotted lines in (b) and (c) are guides for the eye.

Aiming at the design of efficient switchable optical filters, it should be noted that it is also important to bring the maximum transmission contrast as close as possible to the extinction efficiency of the solid Bi nanosphere at the corresponding photon energy. The switching efficiency of a nanosphere can be quantified by the [maximum transmission contrast/$Q_{Extinction\ Solid\ Bi}$] ratio. A perfect switch is achieved when this ratio reaches unity, shown as dashed line in figure 3c. From this figure, it is seen that the best switching efficiency is achieved for D ≤ 30 nm, where the nanosphere [maximum transmission contrast/$Q_{Extinction\ Solid\ Bi}$] ratio seems almost diameter-independent (the points for D → 0, 20 nm and 30 nm can be fitted with a linear function).

Furthermore, to achieve high switching efficiency in a simple metamaterial based on a two-dimensional assembly of Bi nanospheres, $Q_{Extinction\ Solid\ Bi}$ should be as high as possible (and in any case superior to 1). This requirement is crucial to ensure a full extinction of the incoming light by a two-dimensional assembly of solid nanospheres with a reasonable surface coverage.

Taking into account all the previously stated requirements, excellent candidates for building a simple metamaterial with high switching efficiency are the nanospheres with D = 30 nm. They present high [maximum transmission contrast/$Q_{Extinction\ Solid\ Bi}$] ratio and $Q_{Extinction\ Solid\ Bi}$ values. Their transmission contrast takes a maximum value of 0.9 at the near UV photon energy of 3.3 eV, at which the extinction efficiency of the solid Bi nanosphere is 2.8.

## 4. Application: Simple Bi Nanosphere-based Metamaterial for High-Contrast Switchable Ultraviolet Meta-Filters

The optical properties of the Bi nanospheres embedded in a usual dielectric matrix reported in the last section make them interesting building blocks for the design of simple metamaterials for switchable optical filtering applications in the near UV region. The metamaterial structure that we consider in the following consists of a two-dimensional assembly of Bi nanospheres embedded in the same dielectric matrix.

As explained in the previous section, the nanospheres with D = 30 nm are the best candidates for switchable optical filtering applications in the near UV. The maximum of the extinction efficiency $Q_{Extinction}$ of the solid Bi nanosphere is 3.1, at a photon energy around 3.6 eV. Such a value means that the nanosphere would filter a beam in the transmitted direction as a fully opaque flat filter 3.1 times the nanosphere projected area. With such a value, assuming that the transmittance T of a light beam at normal incidence by a two-dimensional assembly of nanospheres can be estimated as T = 1-Cov.$Q_{Extinction}$, Cov being the (relative) coverage of the surface by the nanospheres, a 0% transmittance at 3.6 eV could be achieved with a 32% surface coverage, as shown in figure 4.

The transmittance of the two-dimensional assembly of nanospheres upon Bi melting (the nanosphere shape, size and organization being preserved) is calculated using the same formula as above, with the $Q_{Extinction}$ of a 30 nm liquid Bi nanosphere (figure 2b) and the same surface coverage (32%). Figure 4 shows an almost total extinction (transmission of 2%) of near ultraviolet light at 3.45 eV (359 nm) for the Bi nanospheres in the solid state, whereas upon phase transition the resulting liquid Bi nanospheres show a transmission of 30%. This shows that a proper control of the nanosphere diameter and organization in a simple tailor-made metamaterial already makes possible to achieve a strong liquid-solid *transmittance contrast* (one state showing a near 0% transmittance and the other one a 30% transmittance) that cannot be achieved with Bi thin films. The obtained *transmittance contrast* values are fully suitable for efficient switchable near UV filtering applications.

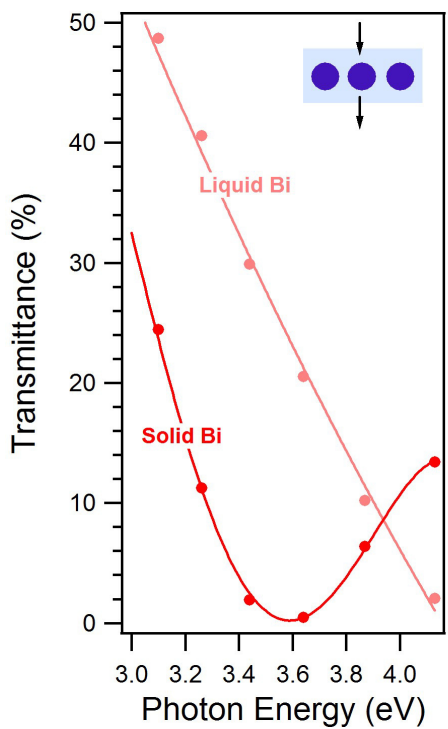

Figure 4 Estimated transmittance of a two-dimensional assembly of solid (dark red dots) and liquid (light red dots) Bi nanospheres embedded in a-$Al_2O_3$, with D = 30 nm and a coverage of 32%. The lines are a guide for the eye.

## 5. Conclusion

Bi nanospheres embedded in an usual and robust dielectric matrix such as a-$Al_2O_3$ include both the "resonator" and "switch" functionalities. This is made possible by the excitation of localized optical resonances that show a change in their nature (from polaritonic to plasmonic) and spectral features (especially their peak photon energy) at the solid-liquid transition of Bi, in relation with the different dielectric functions of solid and liquid Bi. Dipolar resonances with a high peak optical extinction in the near UV region are obtained for nanospheres with 20 nm ≤ D ≤ 50 nm. For a given diameter in this range, the dipolar resonances of a solid and liquid Bi nanosphere show markedly different peak photon energies. Therefore a strong contrast between their extinction efficiencies can be achieved at selected photon energies in the near UV region.

As a consequence, we have proposed a simple metamaterial design, consisting of a two-dimensional assembly of embedded Bi nanospheres that would show a near 0% to 30% optical transmittance switching at the photon energy of 3.45 eV (359 nm) upon reversible solid-liquid transition of Bi (the nanosphere size, shape and organization being kept unchanged by the embedding matrix). These results show that Bi nanostructures markedly subwavelength in size (D ~ λ/10) are powerful candidates for the design of reconfigurable metamaterials for integrated switchable optical filtering solutions in the near UV region. As they include both the "resonator" and "switch" functionalities, such nanostructures are ideal for achieving a local reconfigurability of the metamaterial, down to the nanoscale.

Therefore, this work reveals the importance of Bi in the search for alternative plasmonic materials, beyond noble metals.[18-20,29,33] It appeals at the fabrication of metamaterials based on Bi nanospheres, and to the design of complex metamaterials that will be built from a broader range of embedded Bi nanostructures (nanorods, nanocylinders…). The control in their size, morphology and distribution within the matrix will allow a broad tuning the metamaterials optical response. Furthermore, the comprehensive study of the optical absorption, scattering and extinction properties of Bi nanospheres that we have reported will be useful for applications involving the interaction of light with Bi nanostructures, such as catalysis[32, 41-42], sensing[43] or optoelectronics.[44]

## Acknowledgments

This research received funding from the European Community Seven Framework Programme (FP7-NMP-2010-EU-MEXICO) under Grant Agreement no. 263878. We also acknowledge funding from the Spanish Ministry for Economy and Competitiveness through AMALIE TEC 2012-38901-C02-01 project. We thank Dr Toney Teddy Fernandez (Politecnico di Milano) for valuable advice during the manuscript preparation.

# References

1 Zheludev, N.I. A roadmap for metamaterials, *Optics and Photonic News.* March **2011**, *31*.

2 Gholipour, B.; Zhang, J.; MacDonald, K.F.; Hewak, D.W.; Zheludev, N.I. An all-optical, non-volatile, bidirectional, phase-change meta-switch. *Adv. Mater.* **2013**, *25*, 3050-3054.

3 Vivekchand, S. R. C.; Engel, C. J.; Lubin, S. M.; Blaber, M. G.; Zhou, W.; Suh, J. Y.; Schatz, G. C.; Odom, T. W.; Liquid Plasmonics: Manipulating Surface Plasmon Polaritons via Phase Transitions. *Nano Lett.* **2012**, *12*, 4324-4328.

4 Yang, A.; Hoang, T.B.; Dridi, M.; Deeb, C.; Mikkelsen, M.H.; Schatz, G.C.; Odom, T.W. Real-time tunable lasing from plasmonic nanocavity arrays, *Nat. Comm.* **2015**, *6*, 6939, 1-7.

5 Fafarman, A.T.; Hong, S.-H.; Caglayan, H.; Ye, X.; Diroll, B.T.; Paik, T.; Engheta, N.; Murray, C.N.; Kagan, C.R.; Chemically tailored dielectric-to-metal transition for the design of metamaterials from nanoimprinted colloidal nanocrystals, *Nano Lett.* **2013**, *13*, 350-357.

6 Tittl, A.; Michel, A.-K.U.; Schäferling, M.; Yin, X.; Gholipour, b.; Cui, L.; Wuttig, M.; Taubner, T.; Neubrech, F.; Giessen, H.; A switchable mid-infrared plasmonic perfect absorber with multispectral thermal imaging capability, *Adv. Mater.* **2015**, *02023*, 4597-4603.

7 Lei, D.Y.; Appavoo, K.; Ligmajer, F.; Sonnefraud, Y.; Haglund Jr., R.F.; Maier, S.; Optically-triggered nanoscale memory effect in a hybrid plasmonic-phase changing nanostructure, *ACS Photonics* **2015**, *2*, 1306-1313.

8 Earl, S.K.; James, T.D.; Davis, T.J.; McCallum, J.C.; Marvel, R.E.; Haglund Jr., R.F.; Roberts, A.; Tunable optical antennas enabled by the phase transition in vanadium dioxide, *Opt. Expr.* **2013**, *21*, 27503-27508.

9 Chen, Y.G.; Kao, T.S.; Ng, B.; Li, X.; Luo, X.G.; Luk'yanchuk, B.; Maier, S.; Hong, M.H.; Hybrid phase-change plasmonic crystals for active tuning of lattice resonances. *Opt. Expr.* **2013**, *21*, 13691-13698.

10 Kaplan, G.; Aydin, K.; Scheuer, J.; Dynamically controlled plasmonic nano-antenna phased array utilizing vanadium dioxide, *Opt. Mater. Expr.* **2015**, *5*, 2513-2524.

11 Cao, T.; Zheng, G.; Wang, S.; Wei. C.; Ultrafast beam steering using gradient Au-$Ge_2Sb_2Te_5$-Au plasmonic resonators, *Opt. Expr.* **2015**, *23*, 18029-18039.

12 Markov, P.; Appavoo, K.; Haglund, R.F.; Weiss, S.M.; Hybrid Si-$VO_2$-Au optical modulator based on near-field coupling, *Opt. Expr.* **2015**, *23*, 6878-6887.

13 Zou, L.; Cryan, M.; Klemm, M.; Phase change material based tunable reflectarray for free-space optical inter-intra chip interconnects, *Opt. Expr*. **2014**, *22*, 24142-24148.

14 Matsui, H.; Ho, Y.-L.; Kanki, T.; Tanaka, H.; Delaunay, J.-J.; Tabata, H.; Mid-infrared plasmonic resonances in 2D $VO_2$ Nanosquare arrays, *Adv. Optical Mater.* **2015**, published online.

15 Kooij, E.S.; Ahmed, W.; Zandvliet, H. J. W; Poelsema B. Localized Plasmon In Noble Metal Nanospheroids. *J. Phys. Chem. C.* **2011**, *115*, 10321-10332.

16 Toudert, J.; Babonneau, D.; Camelio, S.; Girardeau, T.; Yubero, F.; Espinós, J.P.; Gonzalez-Elipe, A.R.; Using ion beams to tune the nanostructure and optical response of co-deposited Ag:BN thin films, *J. Phys. D: Appl. Phys.* **2007**, *40*, 4614-4620.

17 Toudert, J; Serna, R and Jiménez de Castro, M. Exploring the optical potential of nano-Bismuth: Tunable surface plasmon resonances in the near ultraviolet to near infrared range. *J. Phys. Chem. C.* **2012**, *116*, 20530-20539.

18 Naik, G.; Shalaev, V.M.; Boltasseva, A.; Alternative plasmonic materials: beyond gold and silver. *Adv. Mater.* **2013**, *25*, 3264-3294.

19 West, P.R.; Ishii, S.; Naik, G.V.; Emani, N.K.; Shalaev, V.; Boltasseva, A.; Searching for better plasmonic materials. *Laser & Photonic Reviews* **2010**, *4*, 795-808.

20 Sanz, J.M.; Ortiz, D.; Alcaraz de la Osa, R.; Saiz, J.M.; González, F.; Brown, A.S.; Losurdo, M.; Everitt, H.O.; Moreno, F.; UV plasmonic behavior of various metal nanoparticles in the near- and far-field regimes: geometry and substrate effects. *J. Phys. Chem. C* **2013**, *117*, 19606-19615.

21 Maidecchi, G.; Gonella, G.; Zaccaria, R.P., Moroni, R.; Anghinolfi, L.; Giglia, L.; Nannarone, S.; Mattera, L.; Dai, H.L.; Canepa, M.; Bisio, F.; Deep Ultraviolet Plasmon Resonance in Aluminum Nanoparticle Arrays, *ACS Nano*, **2013**, *7*, 5834–5841.

22 Ross, M. B., Schatz, G.C.; Aluminium and indium plasmonic nanoantennas in the ultraviolet, *J. Phys. Chem. C*, **2014**, *118*, 12506–12514

23 Watson, A.M.; Zhang, X.; Alcaraz de la Osa, R.; Sanz, J.M.; González, F.; Moreno, F.; Finkelstein, F.; Liu, J.; Everitt, H.O.; Rhodium nanoparticles for ultraviolet plasmonics. *Nano Lett.*, **2015**, *15*, 1095-1100.

24 MacDonald, K.F.; Zheludev, N.I.; Active plasmonics: current status, *Laser Photonics Rev.*, **2010**, *4*, 562-567.

25 Soares, B.F.; MacDonald, K.F.; Fedotov, V.; Zheludev, N.I.; Light-induced switching between structural forms with different optical properties in a single gallium nanoparticulate, *Nano Lett.*, **2005**, *5*, 2104-2107.

26 Jimenez de Castro, F; Cabello, F; Toudert, J; Serna, R; and Haro-Poniatowski, E. Potential of bismuth nanoparticles embedded in a glass matrix for spectral-selective thermo-optical devices. *Appl. Phys. Lett.* **2014,** *105*, 113102, 1-5.

27 Knight, M.W. ; Coenen, T. ; Yang, Y. ; Brenny, B. J. M.; Losurdo, M.; Brown, A. S.; Everitt, H. O.; Polman, A.; Gallium plasmonics: deep subwavelength spectroscopic imaging of single and interacting gallium nanoparticles, *ACS Nano* **2015**, *9*, 2049-2060.

28 Parravicini, G.B.; Stella, A.; Ghigna, P.; Spinolo, G.; Migliori, A.; d'Acapito, F.; Kofman, R.; Extreme undercooling (down to 90K) of liquid metal nanoparticles, *Appl. Phys. Lett.* **2006**, *89*, 033123, 1-3.

29 Blaber, M.G.; Arnold, M.D.; Ford, M.J.; A review of the optical properties of alloys and intermetallics for plasmonics, *J. Phys.: Condens. Matter* **2010**, *22*, 143201, 1-15.

30 Inagaki, T.; Arakawa, E.T.; Cathers, A.R.; Glastad, K.A.; Optical properties of liquid Bi and Pb between 0.6 and 3.7 eV, Phys. Rev. B 1982, 25, 6130-6138.

31 Toudert, J. Spectroscopic ellipsometry for active nano- and meta- materials, *Nanotechnology Reviews* **2014**, *3*, 223-245.

32 Wang, Z. et al.; Investigation of the optical and photocatalytic properties of bismuth nanospheres prepared by a facile thermolysis method, *J. Phys. Chem. C* **2014**, *118*, 1155-1160.

33 McMahon, J.M.; Schatz, G.C.; Gray, S.K; Plasmonics in the ultraviolet with the poor metals Al, Ga, In, Sn, Tl, Pb, and Bi, *Phys. Chem. Chem. Phys.* **2013**, *15*, 5415-5423.

34 Evanoff, D.D. Jr, Chumanov, G. Size-controlled synthesis of nanoparticles. 2. Measurement of extinction, scattering and absorption cross-sections, *J. Phys. Chem. B* **2004**, *108*, 13957-13962.

35 Cuadrado, A; Alda, J; González, F. J. Multiphysics simulation for the optimization of optical nanoantennas working as distributed bolometers in the infrared. *J. Nanophotonics* **2013,** *7*, 073093, 1-15.

36 Karamehmedovic, M.; Schuh, R.; Schmidt, V.; Wriedt, T.; Matyssek, C.; Hergert, W.; Stalmashonak, A.; Seifert, C.; Stranik, O.; Comparison of numerical methods in near-field computation for metallic nanoparticles. *Opt. Expr.* **2011**, *19*, 8939-8953.

37 Silva-López; Cuadrado, A; Llombart, N; Alda, J. Antenna array connections for efficient performance of distributed microbolometers in the IR. *Opt. Expr.* **2013**, *21*, 10867-10877.

38 Khoury, C.G.; Norton, S.J.; Vo-Dinh, T.; Plasmonics of 3-D nanoshell dimers using multipole expansion and finite element method. *ACS Nano* **2009**, *3*, 2776-2788.

39 Cuadrado, A; Silva-López, M; González, F. J; Alda, J. Robustness of antenna-coupled distributed bolometers. *Opt. Lett*, **2013**, *38*, 3784-3787.

40 Kolwas, K; Derkochova, A; Shopa, M. Size characteristics of surface plasmons and their manisfestation in scattering properties of metal particles. *J.Quant. Spectrosc. Radat. Transfer*. **2009**, *110*, 1490-1501.

41 Dong, F.; Xiong, T.; Sun, Y.; Zhao, Z.; Zhou, y.; Feng, X.; Wu, Z.; A semimetal bismuth element as a direct plasmonic photocatalyst, *Chem. Commun.* **2014**, *50*, 10386-10389.

42 Kouamé, N.A.; Alaoui, O.T.; Herissan, A.; Larios. E.; José-Yacaman, M.; Etcheberry, A.; Colbeau-Justin, C.; Remita, H.; Visible light-induced photocatalytic activity of modified titanium(iv) oxide with zero-valent bismuth clusters, *New J. Chem.* **2015**, *39*, 2316-2322.

43 Bazakutsa, A.P.; Golant, K.M.; Near-infrared luminescence of bismuth in fluorine-doped-core silica fibres, *Optics Expr.* **2015**, *23*, 3818-3830.

44 Tian, Y.; Jiang, L.; Deng, Y.; Deng, S.; Zhang, G.; Zhang, X.; Bi-nanorod/Si-nanodot hybrid structure: surface dewetting induced growth and its tunable surface plasmon resonance, *Opt. Mater. Expr.* **2015**, *5*, 1655-2666.